\documentclass[a4paper,UKenglish]{lipics-v2016}

\usepackage{microtype}

\usepackage{rotating}
\usepackage{graphics}
\usepackage{latexsym}
\usepackage[dvips]{epsfig}
\usepackage{amssymb}
\usepackage{graphicx}
\usepackage{epstopdf}
\usepackage{url}
\usepackage{algorithm}
\usepackage{algorithmic}

\newcommand\nc{\newcommand}
\nc{\lem}[2]{\begin{lemma}\label{lem:#1} #2 \end{lemma}}
\nc{\thm}[2]{\begin{theorem}\label{thm:#1} #2\end{theorem}}
\nc{\eqn}[2]{\begin{eqnarray}\label{eqn:#1} #2 \end{eqnarray}}
\nc{\refd}[1]{Definition~\ref{def:#1}}
\nc{\reff}[1]{Fig.~\ref{fig:#1}}
\nc{\refl}[1]{Lemma~\ref{lem:#1}}
\nc{\refp}[1]{Proposition~\ref{prp:#1}}
\nc{\reft}[1]{Theorem~\ref{thm:#1}} \nc{\refe}[1]{(\ref{eqn:#1})}
\nc{\reftb}[1]{Table~\ref{tab:#1}}

\nc{\reffc}[1]{Fact~\ref{fact:#1}}

\def\Z{\mathbb{Z}}

\nc{\pf}[1]{ \noindent \emph{Proof.} #1
 \hfill \qed\par}

\long\def\invis#1{}

\title{A Linear-time Algorithm  for Integral Multiterminal Flows in Trees}

\titlerunning{Integral Multiterminal Flows in Trees} 

\author[1]{Mingyu Xiao}
\author[2]{Hiroshi Nagamochi}
\affil[1]{School of Computer Science and Engineering,
University of Electronic Science and Technology of China, Chengdu, China\\
  \texttt{myxiao@gmail.com}}
\affil[2]{Department of Applied Mathematics and Physics,
  Graduate School of Informatics, Kyoto University, Japan\\
  \texttt{nag@amp.i.kyoto-u.ac.jp}}
\authorrunning{M. Xiao and H. Nagamochi} 

\Copyright{Mingyu Xiao and Hiroshi Nagamochi}


\EventEditors{Seok-Hee Hong}
\EventNoEds{1}
\EventLongTitle{27th International Symposium Algorithms and Computation
(ISAAC 2016)}
\EventShortTitle{ISAAC 2016}
\EventAcronym{ISAAC}
\EventYear{2016}
\EventDate{December 12--14, 2016}
\EventLocation{Sydney, Australia}
\EventLogo{}
\SeriesVolume{64}
\ArticleNo{[42]}

\begin{document}

\maketitle

\begin{abstract}
In this paper, we study
the problem of finding an integral multiflow which maximizes
the sum of flow values between every two terminals
in  an undirected tree with a nonnegative integer edge capacity
and a set of terminals.
In general, it is known that the flow value of an integral multiflow is bounded by
the cut value of a cut-system which consists of disjoint subsets each of which
contains exactly one terminal or has an odd cut value,
and there exists a pair of an integral multiflow and a cut-system whose flow value and cut value
are equal; i.e., a pair of a maximum integral  multiflow and a minimum cut.
In this paper, we propose an $O(n)$-time algorithm that finds such a pair of an integral multiflow and a cut-system
in a given tree instance with $n$ vertices.
This improves the best previous results by a factor of $\Omega (n)$.
Regarding a given tree in an instance as a rooted tree,
we define $O(n)$ rooted tree instances taking each vertex as a root,
and establish a recursive formula on maximum integral multiflow values of these instances
to design
a dynamic programming   that computes the maximum integral multiflow values of
all $O(n)$ rooted instances in linear time.
We can prove that the algorithm implicitly maintains
a cut-system so that not only a maximum integral multiflow but also a minimum cut-system
can be constructed in linear time for any rooted instance whenever it is necessary.
The resulting algorithm is rather compact and succinct.

\end{abstract}

\section{Introduction}

The min-cut max-flow theorem by Ford and Fulkerson \cite{FordFulkerson:flow} is
one of the most important theorems in graph theory.
It catches a min-max relation between two fundamental graph problems.
This theorem leads to many effective algorithms and much theory for flow problems
as well as graph cut problems.
Due to the great applications of it, researchers have interests to seek
more similar min-max formulas in various kinds of flow and cut problems. In this paper, we
consider the \emph{maximum multiterminal flow problem},
a generalization of the basic maximum flow problem.

In the maximum flow problem, we are given two terminals (source and sink)
and asked to find a maximum flow between the two terminals.
A natural generalization of the maximum flow problem is
the famous \emph{maximum multicommodity flow problem}, in which,
a list of pairs of source and sink for the commodities is given and
the objective is to maximize the sum of the simultaneous flows in all the source-sink pairs
subject to the standard capacity and flow conservation requirements.
The maximum multiterminal flow problem is one of the most important special cases
of the maximum multicommodity flow problem. In it, a set $T$ of  more than one terminal
 is given and the list of source-sink pairs is given by all pairs of terminals in $T$.
 The extensions of the maximum flow problem have been extensively studied in the history.
 Readers are referred to a survey~\cite{Costa:Survey}.

A dual problem of the maximum multiterminal flow problem is
the \emph{minimum multiterminal cut problem}, in which
we are asked to find a minimum set of edges whose removal disconnects each pair of terminals in the graph.
The minimum multiterminal cut problem is a generalization of the minimum cut problem.
When there are only two terminals, the min-cut max-flow theorem shows
that the value of the maximum flow equals to the value of the minimum cut in the graph.
However, when there are more than two terminals, the equivalence may not hold.
Consider a star with three leaves. Each leaf is a terminal and each of the three edges has capacity $1$.
The flow value of a maximum multiterminal flow is $1.5$
(a flow of size $0.5$ routed between every pair  of the three terminal pairs),
whereas the size of a minimum multiterminal cut is $2$.
In fact, Cunningham~\cite{Cunningham:multiterminal} has proved a min-max theory for the pair of problems:
The size of a minimum multiterminal cut is at most $(2-2/|T|)$  times
of the flow value of a maximum multiterminal flow.
A similar min-max theory for the maximum multicommodity flow problem
and its dual problem is presented in \cite{FlowMulticut:approximateTheorems}.

In the maximum multiterminal flow problem, each edge is assigned a nonnegative capacity
and a flow  routed between a terminal pair is allowed to take any feasible fraction,
whereas in the \emph{integral multiterminal flow problem}, a flow is allowed
to take a nonnegative integer and we are asked to find a maximum flow under this restriction.
Clearly, we can simply assume that all edge capacities of the integral multiterminal flow problem
are nonnegative integers.
The integral multiterminal flow problem is different from the maximum multiterminal flow problem.
 We can see in the above example,
the flow value of a maximum integral multiterminal flow is $1$.
The special case of the integral multiterminal flow problem
where all edges have unit capacities is also known as the \emph{$T$-path problem},
in which we are asked to find the maximum number of edge-disjointed paths
between different terminal pairs.

In this paper, we study the
maximum multiterminal flow problem in trees
and give linear-time algorithms for both fractional and integer versions,
which improve the best previous algorithms by a factor of $\Omega (n)$~\cite{Costa:tree}.
Note that the maximum (integral) multicommodity flow problem in trees is NP-hard
and there is a $\frac{1}{2}$-approximation algorithm for it~\cite{Garg:multicutinTrees}.

The rest of the paper is organized as follows.
 Section~\ref{pre1} introduces basic notations
on flows and cuts, and reviews important min-max theorems for
fractional and integer versions of maximum multiterminal flow problem.
 Section~\ref{sec:tree-instance} discusses
 instances with rooted trees, and introduces notations   necessary to build
 a dynamic programming method over the set of $O(n)$ instances of rooted subtrees
 of a given instance.
 Informally  ``a blocking flow'' in a rooted tree instance is defined to be a
    flow in the tree currently pushing maximal flows among terminals except for the terminal
    designated as the root.
  Section~\ref{sec:basic} shows several properties of  blocking flows,
    and presents a representation of flow values of blocking flows.
 Section~\ref{sec:main_lemma} provides a main technical lemma
 that tells how to compute the representation of flow values of blocking flows
 and how to construct a maximum flow from the representations.
 Based on the lemma, Section~\ref{sec:algorithm} gives a description of a linear-time algorithm
    for computing the  representations of flow values of blocking flows
    and constructing a maximum flow from the representations.
Finally Section~\ref{conclusions} makes some concluding remarks.

\section{Preliminaries}\label{pre1}

 This section  introduces basic notations
on flows and cuts, and reviews important min-max theorems for
fractional and integer versions of maximum multiterminal flow problem.
Let $\Re^+$ denote the set of nonnegative reals,
and $\Z^+$ denote the set of nonnegative integers.

\medskip

\noindent\textbf{Graphs and Instances}\\
We may denote by $V(G)$ and $E(G)$ the sets of vertices and edges
of an undirected graph $G$, respectively.
Let $G=(V,E)$ denote a simple undirected graph with
a vertex set $V$ and an edge set $E$,
and let $n$ and $m$ denote the number of vertices and edges in a given graph.
Let   $X \subseteq V$ be a subset of vertices in $G$.
Let  $E(X)$ denote  the set of edges with one end-vertex in $X$ and the other in $V-X$,
where $E(\{v\})$ for a vertex $v\in V$ is denoted by   $E(v)$.
Let $G-X$ denote  the graph obtained  from $G$ by removing
 the vertices in $X$ together with the edges in $\cup_{v\in X}E(v)$.
For a vertex subset $T$, let $\mathcal{P}(T)$ be the set of all paths $P_{t,t'}$
with end-vertices $t,t'\in T$ with $t\neq t'$.

An instance $I$ of a maximum flow problem consists
of a graph $G$, a set $T$ of vertices called terminals,
and  a capacity function  $c: E \rightarrow \Re^+$.

\medskip

\noindent\textbf{Flows}\\
For a function $h: E \rightarrow \Re^+$,
   $\sum_{e\in E(X)}h(e)$ for a subset  $X \subseteq V$
is denoted by $h(X)$.
A function $f:E\rightarrow \Z^+$ is called
a {\em   flow} in an instance $(G,T,c)$ if there is
a function $g: \mathcal{P}(T) \to \Z^+$ such that
\[ f(e)=\sum\{ g(P ) \mid e\in E(P ),~P\in \mathcal{P}(T)\}~\mbox{ for all edges $e\in E$,}\]
 where $g(P)$ is the flow value sent along path $P$, and
 such a function $g$ is called a {\em decomposition} of a flow $f$.
A  flow $f$ is called {\em integer} if it admits a   decomposition $g$
such that $g(P )\in \Z^+$ for all paths $P\in \mathcal{P}(T)$
(note that $f$ may not be integer even if $f(e)\in \Z^+$ for all edges $e\in E$).

A flow $f$ is called {\em feasible} if
 $f(e)\leq c(e)$ for all edges $e\in E$.
The {\em flow value} $\alpha(f)$ is defined to be ${\frac{1}{2}}\sum_{t\in T}f(\{t\})$,
and a feasible flow $f$ that maximizes $\alpha(f)$ is called {\em maximum}.

\medskip

\noindent\textbf{Cut-Systems}\\
A subset $X$ of vertices is called a {\em terminal set} (or a {\em $t$-set})
if $X\cap T=\{t\}$ and $X$ induces a connected subgraph from $G$.
A {\em cut-system}  of $T$
 is defined to be a collection $\mathcal{X}$ of disjoint $|T|$ terminal sets
 $X_t$, $t\in T$,
 where  $\mathcal{X}$ is not required to be a partition of $V$.
For  a cut-system  $\mathcal{X}$ of $T$,
let $\gamma(\mathcal{X})=\sum_{X\in \mathcal{X}}c(X)$.
For any pair of a feasible  flow $f$ and a cut-system  $\mathcal{X}$ of $T$
in $(G,T,c)$, it holds
\begin{equation} \alpha(f)\leq \frac{1}{2} \gamma(\mathcal{X}).
\label{real-min-max}
\end{equation}
Cherkasskii~\cite{Cherkasskii}
proved the next result.

\thm{min-max}{
A feasible  flow $f$ in $(G,T,c)$ is maximum
if and only if there is a cut-system $\mathcal{X}$
such that $\alpha(f)=\frac{1}{2} \gamma(\mathcal{X})$.
}\medskip

Ibaraki \emph{et al.} \cite{IKN:inner_euler} proposed an $O(nm\log n)$-time algorithm
for computing a maximum  flow $f$ in a graph $G$ with
$n$ vertices and $m$ edges.
Hagerup \emph{et al.}  \cite{HKNR:treewidthflow} proved
a characterization of the maximum multiterminal flow problem and
gave an $O(\mathrm{ex}(|T|)n)$-time algorithm for the maximum multiterminal flow problem in bounded treewidth graphs,
where $\mathrm{ex}(|T|)$ is an exponential function of the number $|T|$ of terminals.
This algorithm runs in linear time only when $|T|$ is restricted to a constant.

An integer version of the multiterminal flow problem is
defined as follows.
Let  $I=(G=(V,E),T,c)$ have integer capacities
$c(e)\in \Z^+$, $e\in E$.
Recall that an  integral  flow $f$ is a  flow
which can be
decomposed into integer individual flows $g$,
i.e.,  $g:{\cal P}(T)\rightarrow \Z^+$.
An instance $(G,T,c)$ is called {\em inner-eulerian}
if all edge capacities $c(e)$, $e\in E$ are integers
and $c(E(v))$ is an even integer for each non-terminal vertex $v\in V-T$.
It is known that any inner-eulerian instance admits
a pair of a maximum integral  flow $f$ and
a cut-system $\mathcal{X}$ with $\alpha(f)=\frac{1}{2}\gamma(\mathcal{X})$~\cite{Cherkasskii}.
In general, there is no pair of
an integral  flow $f$ and a cut-system $\mathcal{X}$
with $\alpha(f)=\frac{1}{2}\gamma(\mathcal{X})$ even for trees.
We review a min-max theorem on the integer version as follows.

Assume that  $c(e)\in \Z^+$, $e\in E$.
A component $W\subseteq V$ in the graph $G- \cup_{X\in \mathcal{X}}X$
is called an {\em odd set} in  $\mathcal{X}$ if $c(W)$ is odd.
Let $\kappa(\mathcal{X})$ denote the number of odd sets in $G- \cup_{X\in \mathcal{X}}X$.
For each odd set $W$, at least one unit of capacity from $c(W)$ cannot be
used by any  feasible integral  flow $f:E\to \Z^+$.
Hence since each path in $\mathcal{P}(T)$  goes through
edges in $E(X_t)$ of a $t$-set for exactly two terminals $t\in T$,
we see that, for any decomposition $g$ of $f$,
\begin{equation}
 2\alpha(f)=\sum_{P\in \mathcal{P}(T)}g(P) \leq \sum_{X\in \mathcal{X}}c(X) -\kappa(\mathcal{X})= \gamma(\mathcal{X}) - \kappa(\mathcal{X}).
\label{ingeter-min-max}
\end{equation}
Mader \cite{mader1978} proved the next result.

\thm{min-max_integer}{
A feasible  integral flow $f$ in $(G,T,c)$ is maximum
if and only if there is a cut-system $\mathcal{X}$
such that $\alpha(f)=\frac{1}{2} [\gamma(\mathcal{X})-\kappa(\mathcal{X})]$.
}

For trees with $n$ vertices, an $O(n^2)$-time algorithm
for computing a maximum integral  flow $f$ is proposed
\cite{Costa:tree}, while no strongly-polynomial time
algorithm is known to general graphs (e.g., see \cite{Costa:Survey}).

\section{Tree Instances}\label{sec:tree-instance}

In the rest of this paper, we assume that a given instance $I=(G,T,c)$
consists of a tree $G=(V,E)$, a terminal set $T$ and an integer capacity
$c(e)\in \Z^+$ for each $e\in E$.
We simply call an integral flow a {\em flow}.

This section discusses
 instances with rooted trees, and introduces notations   necessary to build
 a dynamic programming method over the set of $O(n)$ instances of rooted subtrees
 of a given instance.

\begin{figure}[htbp]
\begin{center}
\includegraphics[scale=0.45]{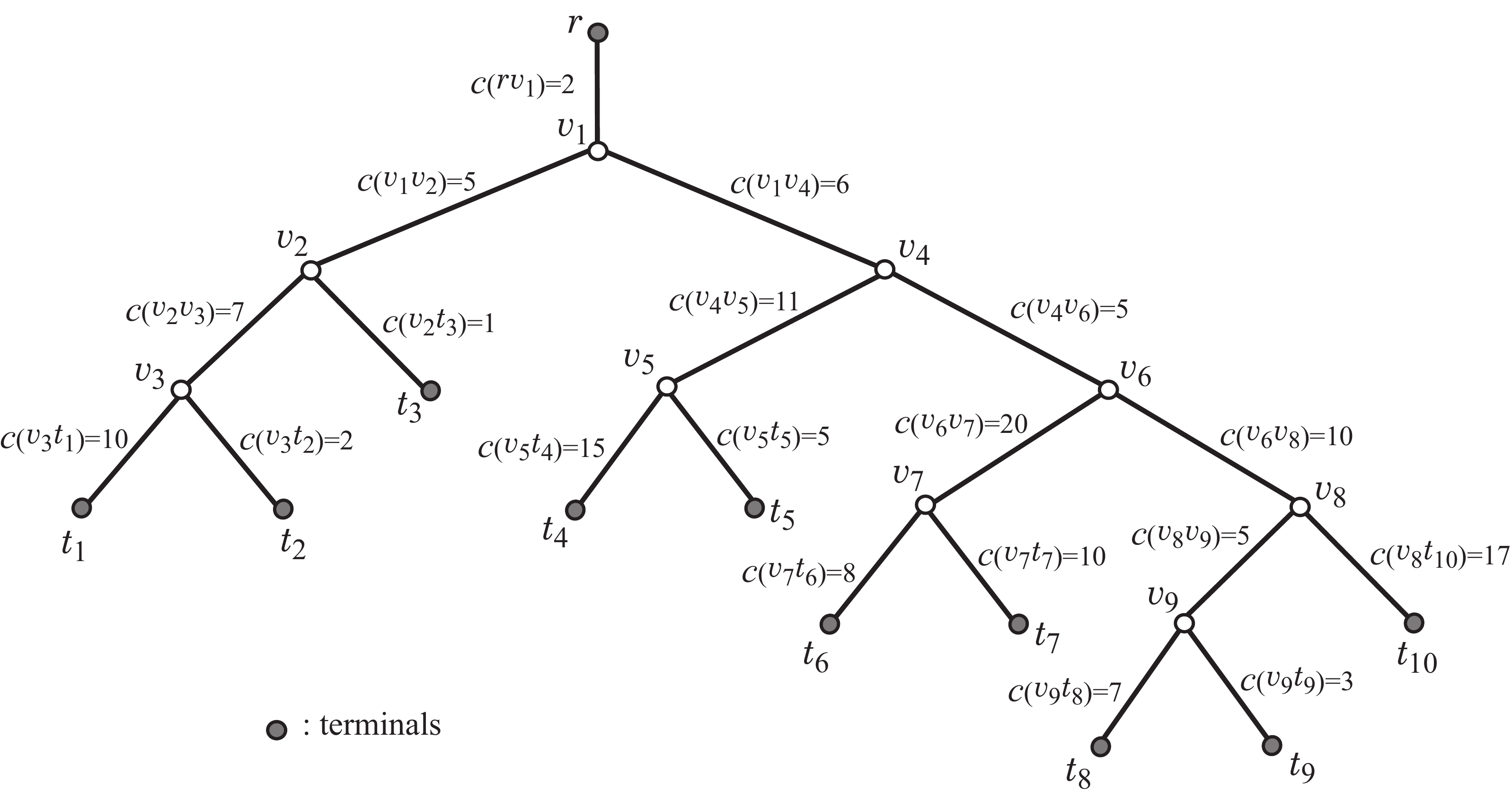}
\end{center}
\caption{An example of a tree instance $I=(G,T,c)$ such that
the degree of each internal vertex is 3 and all capacities are positive integers,
where terminal $r$ is chosen as the root.}
\label{fig:instance_integer1}
\end{figure}

If a vertex $v\in T$ is not a leaf of $G$, i.e., $v$ is of degree $d\geq 2$,
then we can split the instance at the cut-vertex $v$ into $d$ instances, and
it suffices to find a maximum flow in each of these instances.
Also we can split a vertex $v\in V-T$ of degree $d\geq4$ into
$d-2$ vertices that induce a tree with edges of capacity sufficiently larger
without losing the feasibility and optimality of the instance.
In the rest of paper, we assume that $T$ is the set of leaves of $G$, and
the degree of each non-leaf is 3, and
$c(e)\geq 1$ for all edges $e\in E$,
 as shown in Fig.~\ref{fig:instance_integer1}.

For a leaf $v\in V$ in $G$, let $e_v$ denote the edge incidenet to $v$.
For two vertices $u,v\in V$, let $P_{u,v}$ denote the
path connecting $u$ and $v$ in the tree $G$.
For a subset $S\subseteq V$ of vertices,
let $\mathcal{P}(S)$ denote the set of all paths $P_{s,s'}$ with $s,s'\in S$.

In a tree instance $(G,T,c)$,
a {\em   flow} admits
a function $g: {{T}\choose {2}} \to \Z^+$ such that
\[ f(e)=\sum\{ g(t,t') \mid e\in E(P_{t,t'}),~t,t'\in T\}~\mbox{ for all edges $e\in E$,}\]
 where $g(t,t')$ is the flow value sent along path $P_{t,t'}$.
  For a flow $f$, a path $P\in \mathcal{P}(T)$ is called a {\em positive-path}
if $f$ admits a decomposition $g$ such that $g(t,t')>0$.

For a path $P$ in $G$,
and an integer $\delta\geq  - \min_{e'\in E} h(e')$ (possibly $\delta<0)$,
the function $h':E\rightarrow \Z^+$ obtained from $h$
by setting $h'(e)=h(e)+\delta$ for all edges $e\in E(P)$
and $h'(e)=h(e)$ for all edges $e\in E-E(P)$ is denoted
by $h+(P,\delta)$.

\medskip

\noindent\textbf{Rooted Tree}\\
Choose a terminal $r\in T$, and regard $G$ as
a  tree rooted at $r$, which defines a parent-child relationship among
the vertices in $G$.
In a rooted tree $G$,  we write an edge $e=uv$ such that $u$ is the parent of $v$
by an ordered pair $(u,v)$.
 For an edge $e=(u,v)$,
any edge $e'=(v,w)$ is called a {\em child-edge} of $e$,
and $e$ is called the {\em parent-edge} of $e'$.

Let $Y$ be
a subset   of vertices in $V-\{r\}$ such that
$Y$ induces a connected subgraph from $G$.
Then
there is exactly one edge $(u,v)\in E(Y)$   such that $v\in  Y$ and
$u$ is the parent of $v$,
and we call the edge $uv$ the {\em parent-edge} of $Y$
while any other edge  in $E(Y)$
is called a {\em child-edge} of $Y$.

For an edge $e=(u,v)\in E$, let
$V_e\subseteq V$ denote the set of vertex $u$ and all the descendants of $v$ including $v$ itself,
 $G_e=(V_e, E_e)$ denote  the graph   induced from $G$ by $V_e$, and
 let $T_e= (T\cap V_e)-\{u\}$, where we remark that $u\not\in T_e$.
Let $I(e)$ denote an instance $(G_e,T_e\cup\{u\}, c)$ induced from $(G,T,c)$ by
the vertex subset $V_e$,   where we remark that $u$ is included
as a terminal in the instance $I(e)$.

\medskip

\begin{figure}[htbp]
\begin{center}
\includegraphics[scale=0.38]{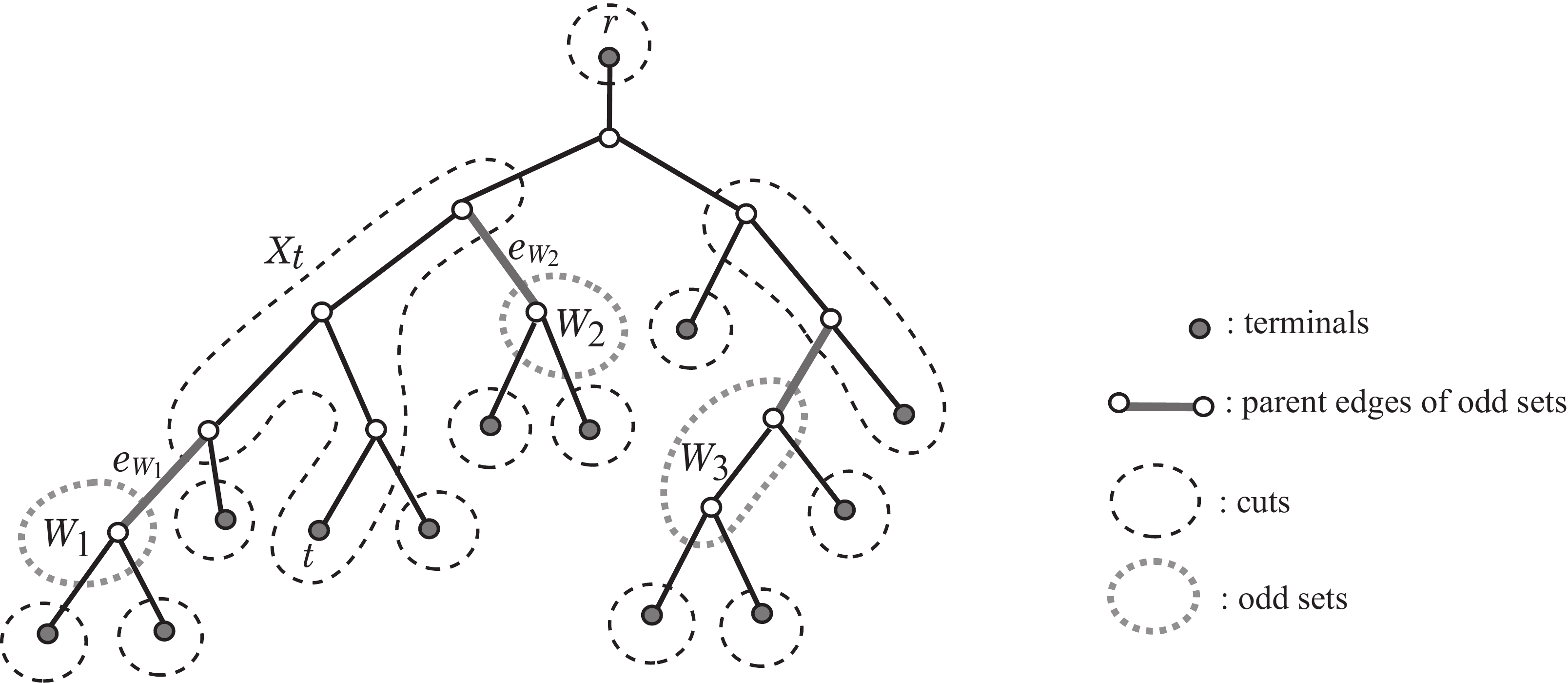}
\end{center}
\caption{Illustration of a cut-system $\mathcal{X}$
and the family $\mathrm{odd}(X_t)=\{W_1,W_2\}$
 for a terminal set $X_t\in \mathcal{X}$. }
\label{fig:odd_set}
\end{figure}

\noindent\textbf{Blocking Flows}\\
Informally  ``a blocking flow'' in a rooted tree instance is defined to be a
    flow in the tree currently pushing maximal flows among terminals except for the terminal
    designated as the root.
Let $\mathcal{X}$ be a cut-system of $T_e$ in $I(e)$ for some edge $e=(u,v)$.
An odd set $W$ in $G_e-\cup_{X\in \mathcal{X}}X$
 is called an {\em odd set} of a terminal set $X\in {\cal X}$
if the parent-edge of $W$ is a child-edge of $X$,
where $u\not\in X$ implies    $r,u\not\in W$.
For each terminal set $X\in \mathcal{X}$,
let $\mathrm{odd}(X)$ denote the family of odd sets of $X$, i.e.,
$W$ of $\mathcal{X}$ whose parent-edge $e_W$ is a child-edge of $X$.
  Fig.~\ref{fig:odd_set} illustrates a cut-system $\mathcal{X}$
  and the family $\mathrm{odd}(X_t)=\{W_1,W_2\}$.

For a function $h:E\to \Re_+$,
let $E[h; k]$ denote the set of edges $e\in E$ such that
$h(e)\geq k$.

Let $f$ be  a feasible flow   of $I(e)$ for an edge $e=(u,v)$.
We call
 a terminal set $X\in \mathcal{X}$ with $t\in X\cap T$
  {\em blocked}  (or {\em blocked by $f$}) if
  \[f(e_t)=f(X)=c(X)-|\mathrm{odd}(X)|,\]
and call $\mathcal{X}$
  {\em blocked}  (or {\em blocked by $f$})
  if all terminal sets in it are  blocked by $f$.

For each vertex $s\in V_e$, we define
  $V_f(s)$   to be the set of vertices $w\in V_e$
   reachable from $s$ 
by a path $P_{s,w'}$ from $s$ to the common ancestor $w'$ of $s$ and $w$
using edges in $E[c-f;1]$ and
by a path $P_{w',w}$ from $w'$ to $w$ using edges in
  $E[c-f;2]$.
In other words, we travel an edge $e'$ upward if $c(e')-f(e')\geq 1$
and downward if $c(e')-f(e')\geq 2$ from $s$ to $w$.
By the definition of $V_f(s)$, we can see that $V_f(s)$ induces a connected subgraph,
the parent-edge $e'$ of $V_f(s)$ satisfies $f(e')=c(e')$,
  and any child-edge $e'$ of $V_f(s)$ satisfies $f(e')\in \{c(e')-1,c(e')\}$.

We call $f$ {\em blocking} if $\{V_f(t) \mid t\in T_e\}$  is   a cut-system  of $T_e$
blocked by $f$.
Let $\Psi(e)$ denote the set of integers $x$ such that
$I(e)$ has a blocking flow $f(e)=x$.

\medskip

\noindent\textbf{Interval Computation}\\
Our dynamic programming approach to compute the maximum flow value
updates the set of flow values of blocking flows recursively.
As it will be shown in Section~\ref{sec:basic},
such a set of flow values always is given by an interval that consists of consecutive odd or even integers,
and we here introduce a special operation on such types of intervals.

 For two  reals $a,b$ with $a\leq b$,
 let $[a,b]$ denote the set of  reals $s$ with $a\leq s\leq b$.

For two integers $k,a\in \Z^+$,
the set $\{a+2i\mid i=0,1,\ldots,k\}$ of consecutive odd or even integers
is denoted by $\langle a, b\rangle$, where $b=2k+a$.
For two sets $A,B\subseteq \Z^+$ of nonnegative integers,
let $A\otimes B$ denote the set of nonnegative integers
$\{a+b-2i\mid  i=0,1,\ldots,\min\{a,b\}\}$ over all
$a\in A$ and $b\in B$.
In particular,
for sets $A_1=\langle a_1,b_1\rangle$ and  $A_2=\langle a_2,b_2\rangle$,
we observe that
\[A_1\otimes A_2 =
  \left\{ \begin{array}{ll}
 \mbox{$\langle 0, b_1+b_2 \rangle$} & \mbox{if $A_1\cap A_2\neq\emptyset$} \\
 \mbox{$\langle 1, b_1+b_2 \rangle$} &
 \mbox{if $a_2\leq b_1 $, $a_1\leq b_2$ and $A_1\cap A_2=\emptyset$} \\
 \mbox{$\langle a_1-b_2, b_1+b_2 \rangle$} & \mbox{if $ b_2<a_1$} \\
 \mbox{$\langle a_2-b_1, b_1+b_2 \rangle$} & \mbox{if $ b_1<a_2$.}
 \end{array} \right.
\]

Given an integer $x\in A_1\otimes A_2$, we can find in $O(1)$ time
three integers $x_i\in \langle a_i, b_i\rangle$, $i=1,2$ and $y\in [0, \min\{x_1,x_2\}]$
 such that
$x=x_1+x_2 -2y$.
To see this, assume that $b_1\leq b_2$ without loss of generality, and let $a'_2$ be the minimum element in $\langle a_2,b_2\rangle$ with $b_1\leq a'_2$, where $a'_2\in \{b_1,b_1-1,a_2\}$.
Observe that
$\{x\in A_1\otimes A_2\mid x\leq b_2-b_1\}=
 \{b_1+x_2-2b_1\mid x_2\in \langle a'_2,b_2\rangle\}$ and
$\{x\in A_1\otimes A_2\mid x> b_2-b_1\}=\{b_1+b_2-2y \mid y=0,1,\ldots,b_1-1\}$.
Hence if $x\leq b_2-b_1$ then let $x_1=y=b_1$ and $x_2=x+b_1$; otherwise
$x_1=b_1$, $x_2=b_2$ and $y=(x-b_1-b_2)/2$.

 \section{Basic Properties on Blocking Flows}  \label{sec:basic}

 This section shows several properties of  blocking flows,
    and presents a representation of flow values of blocking flows.
We first observe two lemmas on some properties of blocking flows.

\lem{integral-blocking-flow}
{Let $f$ be a feasible flow in $I(e)$   for an edge $e\in E$.
\begin{enumerate}
\item[{\rm (i)}]
For a terminal $t\in T_e$, let $X_t$ be a $t$-cut such that $f(X_t)=f(e_t)$
and $P_{s,s'}$ be a  positive-path of $f$ with $s,s'\in T_e\cup\{u\}$.
If $t\in \{s,s'\}$ then $P_{s,s'}$  contains exactly one edge in $E(X_t)$,
 and otherwise
     $P_{s,s'}$ is disjoint with $X_t$.

\item[{\rm (ii)}]
Assume that  $V_f(t)\cap V_f(t')=\emptyset$   for any two $t,t'\in T_e$.
Then $V_f(u)$ is disjoint with $V_f(t)$ of any terminal $t\in T_e$, and the following holds:\\
 {\em (1)} For each edge $e'\in E(V_f(t))$ with $t\in T_e\cup\{u\}$,
 \[
f(e') = \left \{\begin{array}{cl}
    c(e')-1 & \mbox{if $e'$ is the parent-edge of an odd set $W\in \mathrm{odd}(V_f(t))$} \\
      c(e') & \mbox{otherwise.}
   \end{array}
  \right.
 \]
{\rm (2)} $f(V_f(t))=c(V_f(t)) -|\mathrm{odd}(V_f(t))| $ for each $t\in T_e\cup\{u\}$.

\item[{\rm (iii)}]
Flow $f$ is blocking if   $V_f(t)\cap V_f(t')=\emptyset$   for any two $t,t'\in T_e$,
and $f(e_t)=f( V_f(t) )$ for each   $t\in T_e$.

\item[{\rm (iv)}]
When $f$ is blocking,    any edge $e'\in E_e$ with $f(e')=c(e')$ satisfies $c(e')\in \Psi(e')$.

\item[{\rm (v)}]
When $f$ is blocking,
 the parent-edge $e_W$ of any odd set $W\in \mathrm{odd}(V_f(t))$ for a terminal $t\in T_e$
 satisfies $c(e_W)-1\in \Psi(e')$.
\end{enumerate}
}

\pf{
\noindent (i)
Let $g$ be a  decomposition of $f$ such that $g(s,s')>0$ for some terminals $s,s'\in T_e\cup\{u\}$.
Since $X\cap (T_e\cup\{u\})=\{t\}$, we obtain
 $f(e_t)=\sum_{t'\in (T_e-\{t\})\cup\{u\}}g(t,t')  \leq f(X_t)$.
Hence  the positive-path $P_{s,s'}$ with $s\neq t\neq s'$
contains an edge in $E(X_t)$, then $f(e_t)< f(X_t)$ would hold,
contradicting $f(e_t)=f(X_t)$.
 Clearly if $t\in \{s,s'\}$ then $P_{s,s'}$  contains exactly one edge in $E(X_t)$. \\

\noindent (ii)
We see that $V_f(u)$ is disjoint with $V_f(t)$ of any terminal $t\in T_e$,
since the vertices in $V_f(u)$ are spanned with  edges in $E[c-f; 2]$
and the parent-edge of $V_f(t)$ is saturated by $f$.

Let $\hat{e}^t$ denote the parent-edge of $V_f(t)$, $t\in T_e$, where  $V_f(u)$ has no parent-edge in $I(e)$.
By construction, the parent-edge $\hat{e}^t$ of $V_f(t)$ with $t\in T_e$ is saturated by $f$
and any child-edge $e'$ of $V_f(t)$ with $t\in T_e\cup\{u\}$  satisfies $f(e')\in \{c(e'), c(e')-1 \}$.

We prove (1) and (2) by induction on the size $|V_{\hat{e}^t}-V_f(t)|$.
As the base case where $t\in T_e$ is a terminal such that  $V_f(t)$ has no child-edge, i.e., $V_f(t)=\{t\}$,
we see that $\mathrm{odd}(V_f(t))=\emptyset$ and
$f(V_f(t))=f(\{t\})=f(\hat{e}^t)=c(\hat{e}^t)=c(\{t\})=c(V_f(t)) -|\mathrm{odd}(V_f(t))|$
and all edges $e'\in E(V_f(t))$ are saturated by $f$, proving (1) and (2) for such a terminal $t$.

Next let $t$ be a terminal in $T_e\cup\{u\}$ such that
 the properties (1) and (2) are assumed to hold
 for
all $t'$-cuts $V_f(t')$ such that $t'\neq t$ and  $V_f(t')\subseteq V_{\hat{e}^t}$,
as an inductive hypothesis.
Then
the child-edges of any odd set $W\in \mathrm{odd}(V_f(t))$ are saturated
and thereby the parent-edge $e_W$ of $W$ must satisfy $f(e_W)=c(e_W)-1$
since $W$ contains no terminal and $c(W)$ is odd.
Therefore any child-edge of $V_f(t)$ is either the parent-edge of an odd set  $W\in \mathrm{odd}(V_f(t))$,
where $f(e')=c(e')-1$, or the parent-edge of a cut $V_f(t')$, where
  $f(e')=c(e')$ holds, proving (1) for $t$.
  Since $f(\hat{e}^t)=c(\hat{e}^t)$ or $V_f(u)$ has no parent-edge,
  this means  $f(V_f(t))=c(V_f(t))-|\mathrm{odd}(V_f(t))|$,
  proving (1) for the terminal $t$.
  This completes the inductive proof for the properties (1) and (2). \\

\noindent (iii) Assume that  $V_f(t)\cap V_f(t')=\emptyset$   for any two $t,t'\in T_e$,
and $f(e_t)=f( V_f(t) )$ for each   $t\in T_e$.
Then by the result of (ii), we have
  $f(V_f(t))=c(V_f(t))-|\mathrm{odd}(V_f(t))|$ holds for all $t\in T_e$.
  Since $f(e_t)=f( V_f(t) )$ for each   $t\in T_e$, each set $V_f(t)$ with $t\in T_e$ is blocked by $f$,
  and
 the family $\{V_f(t) \mid t\in T_e\}$ is a cut-system blocked by $f$.
 Hence $f$ is blocking.  \\

\noindent (iv)
Let $\mathcal{X}=\{V_f(t) \mid t\in T_e\}$, which is
blocked by $f$ by definition.
Let $e'=(u',v')\in E_e$ satisfy $f(e')=c(e')$,
where $u'$ is the parent of $v'$,
and let $f'$ be  the flow in $I(e')$ induced from $f$ by $V_{e'}$.
To show $f'(e')=c(e')\in \Psi(e')$,
it suffices to prove that $f'$ is blocking, i.e.,
$\{V_{f'}(t) \mid t\in T_{e'}\}$ is a  cut-system of $T_{e'}$ blocked by $f'$.

For each terminal $t\in T_{e'}$, the set $V_f(t)$ includes an ancestor $w$ of $t$
when the path $P_{w,t}$ consists of unsaturated edges, and hence
$u'\not\in V_f(t)$ since $e'$ is saturated by $f$.
This means that
  $\{V_{f'}(t) \mid t\in T_{e'}\}=\{V_{f}(t) \mid t\in T_{e'}\}$,
  which is a  cut-system of $T_{e'}$ blocked by $f'$. \\

\noindent (v)
Let $e_W=(u_W,v_W)$,
where $u_W$ is the parent of $v_W$.
It suffices to show that the flow $f'$ in $I(e_W)$ induced from $f$ by $V_{e_W}$
is a blocking flow in $I(e_W)$, i.e.,
$\{V_{f'}(t) \mid t\in T_{e_W}\}$ is a cut-system of $T_{e_W}$ blocked by $f'$.

Since $e_W$ is the parent-edge of odd set $W$,
all child-edges of $W$ are saturated by $f$
by the result of (ii).
For each terminal $t\in T_{e_W}$, the set $V_f(t)$ includes an ancestor $w$ of $t$
when the path $P_{w,t}$ consists of unsaturated edges.
From these observations, we see  that
  there is no terminal $t\in T_{e_W}$ such that $V_f(t)\cap W\neq\emptyset$, and we have
  $\{V_{f'}(t) \mid t\in T_{e_W}\}=\{V_{f}(t) \mid t\in T_{e_W}\}$,
  which is a  cut-system of $T_{e_W}$ blocked by $f'$.
}

\medskip

The next lemma tells how to obtain a maximum flow and a minimum cut-system in an instance $I(e)$.

\lem{last_integral_flow}
{For an edge $e=(u,v)\in  E$,
let $f$ be a blocking  flow in $I(e)$ such that $f(e)$ is the maximum in $\Psi(e)$.
Then $\mathcal{X}=  \{V_f(t)\mid t\in T_e\cup\{u\}\}$ is
a cut-system in $I(e)$ satisfying
$2\alpha(f)=f(e)+\sum_{t\in T_e }f(e_t) = \gamma(\mathcal{X})-\kappa(\mathcal{X}) $
$($hence $f$ is a maximum flow in $I(e)$ by $($\ref{ingeter-min-max}$))$.
}
\pf{  Since $f$ is a blocking flow  in $I(e)$,
the family $\{ V_f(t) \mid t\in T_e\}$ is a cut-system of $T_e$ blocked by $f$
by definition, and we know  that
 $f(e_t)=f(V_f(t))=c(V_f(t))-|\mathrm{odd}(V_f(t))|$ for all terminals $t\in T_e$.
First we see that $V_f(u)$ is disjoint with $V_f(t)$ of any terminal $t\in T_e$,
since the vertices in $V_f(u)$ are spanned with  edges in $E[c-f; 2]$
and the parent-edge of $V_f(t)$ is saturated by $f$.
By \refl{integral-blocking-flow}(ii), we have $f(V_f(u))=c(V_f(u))-|\mathrm{odd}(V_f(u))|$.

We now show that $f(e)=f(V_f(u))$.
If $f(e)\in\{c(e), c(e)-1\}$, then we have $V_f(u)=\{u\}$
and $f(e)=f(V_f(u))$.
Consider the case where  $c(e)-f(e)\geq 2$.
We claim that any  positive-path  $P_{t_1,t_2}$ for $t_1,t_2\in T_e$
is disjoint with  $V_f(u)$.
Assume indirectly that a positive-path  $P_{t_1,t_2}$
contains a vertex in $V_f(u)$.
Let $w$ be the branch vertex of  $P_{t_1,u}$ and $P_{t_2,u}$.
The function $f':=f+(P_{t_1,t_2},-1)+(P_{t_1,u},1)+(P_{t_2,u},1)$
is a feasible flow in $I(e)$, since
$V_f(u)$ is spanned with   edges in $E[c-f;  2]$.
Since $f'(e')=f(e')$ for all edges $e'\in E-E(P_{u,w})$,
the cut-system  $\mathcal{X}$ is blocked also by
the flow $f'$,   and thereby $f'$ is a blocking flow in $I(e)$
with $f'(e)>f(e)=\max\{x\in \Psi(e)\}$, which contradicts
the definition of $\Psi(e)$.
Hence any  positive-path $P_{t_1,t_2}$ with $t_1,t_2\in T_e$
 is  disjoint with $V_f(u)$.
This proves that $f(e)=f(V_f(u))$ even if $c(e)-f(e)\geq 2$.
It always holds that
$f(e)=f(V_f(u))=c(V_f(u))-|\mathrm{odd}(V_f(u))|$.
Therefore we have
$2\alpha(f)=f(e)+ \sum_{t\in T_e}f(e_t)=
\sum_{t\in T_e}(c(V_f(t))-|\mathrm{odd}(V_f(t))|) +
  c(V_f(u))-|\mathrm{odd}(V_f(u))| = \gamma(\mathcal{X})-\kappa(\mathcal{X}) $, as required.
}


\medskip
We prove that all edges $e\in E$  satisfies the following conditions (a)  and (b)
by an induction of depth of edges.\\

\noindent
{\bf (a)}
   $\Psi(e)$ is given by $\langle a(e), b(e)\rangle$
  with some   integers $a(e)$ and $b(e)$
   such that
\begin{enumerate}
\item[{\rm (i)}]
 For each leaf-edge $e$, it holds $\Psi(e)= \langle a(e)=c(e), b(e)=c(e)\rangle$;

\item[{\rm (ii)}]
 For each non-leaf-edge $e$ with two child-edges $e_1$ and $e_2$,
 it holds
\[ \Psi(e)=\langle a(e), b(e)\rangle= ( (\Psi(e_1)\otimes\Psi(e_2) )  \cap   [0, c(e)]) \cup \{c(e)\}. \]
That is,
for $\langle \tilde{a}(e), \tilde{b}(e)\rangle = \Psi(e_1)\otimes\Psi(e_2)$, where
 $\tilde{b}(e)=b(e_1)+b(e_2)$ and
\begin{equation}\label{eq:update-1}
  \tilde{a}(e) =
  \left\{ \begin{array}{cl}
  0   &
   \mbox{if ``$a(e_2)< b(e_1) $ or $a(e_1)<  b(e_2)$'' and $a(e_1)+a(e_2)$ is  even,} \\
   1    &
    \mbox{if ``$a(e_2)<  b(e_1) $ or $a(e_1)<  b(e_2)$'' and  $a(e_1)+a(e_2)$ is odd,} \\
  a(e_i)-b(e_j)   & \mbox{if $ b(e_j)+2\leq a(e_i)$ with $\{i,j\}=\{1,2\}$,}
 \end{array} \right.
\end{equation}
where  edge $e_1$ (resp., $e_2)$ is called
  {\em dominating} if
    $ b(e_2)+2\leq a(e_1)$
  (resp., $ b(e_1)+2\leq a(e_2)$),
it holds that
\begin{equation}\label{eq:update-2}
 \langle a(e), b(e)\rangle =
  \left\{ \begin{array}{ll}
  \langle \tilde{a}(e),  \tilde{b}(e)  \rangle & \mbox{if $\tilde{b}(e)\leq  c(e)$,} \\
   \langle \tilde{a}(e), c(e) \rangle  &
     \mbox{if $\tilde{a}(e)\leq c(e)< \tilde{b}(e)$ and $\tilde{a}(e)+c(e)$ is  even,} \\
  \langle \tilde{a}(e), c(e)\!-\!1 \rangle  &
     \mbox{if $\tilde{a}(e)\leq c(e)< \tilde{b}(e)$ and $\tilde{a}(e)+c(e)$ is   odd,} \\
 \langle c(e), c(e)\rangle  & \mbox{if $c(e) <  \tilde{a}(e)$.}
 \end{array} \right.
\end{equation}
\end{enumerate}

\noindent
{\bf (b)}  If $e=(u,v)$ has a dominating child-edge  $e'=(v,w)$,  then
   there is a   terminal $t\in T_{e'}$ such that  $g({u,t})\geq a(e)$  holds
for any decomposition $g$ of a blocking flow $f$ to $I(e)$
and
 $P_{v,t}$ consists of dominating edges. \\

\begin{figure}[htbp]
\begin{center}
\includegraphics[scale=0.40]{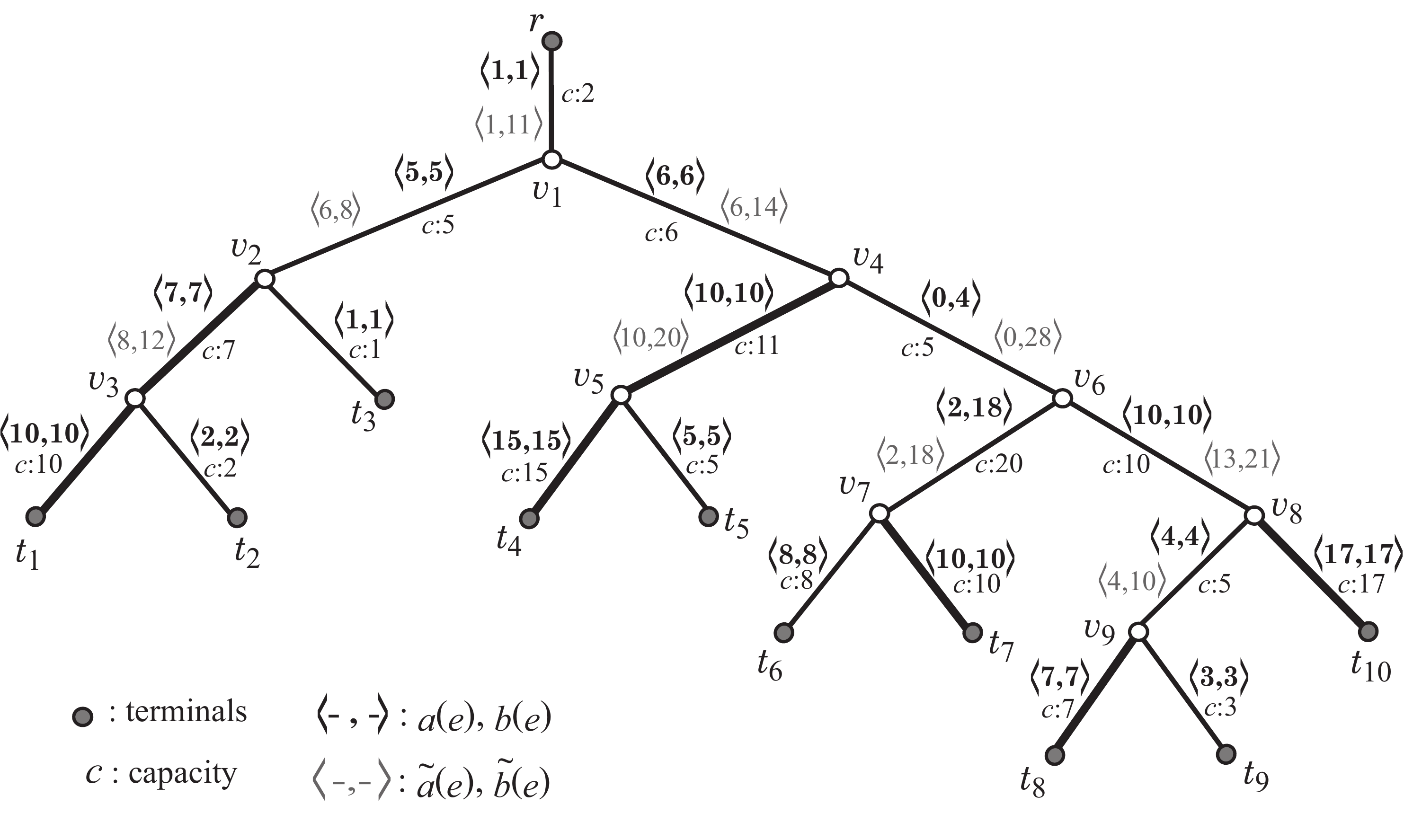}
\end{center}
\caption{A pair of integers $\tilde{a}(e)$ and $\tilde{b}(e)$
in   (\ref{eq:update-1})  and
$\Psi(e)=\langle a(e), b(e)\rangle$ in
  (\ref{eq:update-2}) for each edge $e\in E$
in the instance $I$ in Fig.~\ref{fig:instance_integer1},
where each pair of  $a(e)$ and $b(e)$ is depicted in bold while
that of  $\tilde{a}(e)$ and $\tilde{b}(e)$ in gray.
The dominating edges are depicted in thick lines.}
\label{fig:instance_integer2}
\end{figure}

A path consisting of dominating edges is called a {\em dominating path}.
Fig.~\ref{fig:instance_integer2} shows the pairs $\{\tilde{a}(e),\tilde{b}(e)\}$ and $\{a(e),b(e)\}$
for all edges $e\in E$
in the instance $I$ in Fig.~\ref{fig:instance_integer1} computed according
to (\ref{eq:update-1}) and (\ref{eq:update-2}).

Assuming
that each edge  with depth at least $d$ satisfies   conditions (a)  and (b),
we prove that any edge $e$ with depth $d-1$ satisfies the statements in the next lemma,
which indicates not only conditions (a) and (b) for the edge $e$ but also
how to construct a blocking flow in $I(e)$ from blocking flows in $I(e_1)$ and $I(e_2)$
of the child-edges $e_1$ and $e_2$ of $e$.

 \section{Main Lemma}  \label{sec:main_lemma}

This section  provides a main technical lemma
 that tells how to compute the representation of flow values of blocking flows given by conditions (a)  and (b),
 and how to construct a maximum flow from the representations.

\lem{induction_integral_flow}{
Let $e=(u,v)$ be a non-leaf-edge with depth $d-1~(\geq 1)$.
Assume that all edges with depth at least $d$ satisfy conditions {\rm (a)} and {\rm (b)}.
For the  two children $w_1$ and $w_2$ of $v$,
 let $\langle \tilde{a}, \tilde{b}\rangle
 =\Psi(vw_1)\otimes\Psi(vw_2)=\langle a(vw_1), b(vw_1)\rangle
   \otimes \langle a(vw_2), b(vw_2)\rangle$.

\begin{enumerate}
\item[{\rm (i)}]
For  a blocking flow   of $I(e)$,
if $e\in E(V_f(t))$ for some terminal $t\in T_e$,
then
the path $P_{v,t}$ from $v$ to $t$ is a dominating path,
the path $P_{u,t}$ from $u'$ to $t$ satisfies $g(u,t)\geq c(e)$
for any decomposition $g$ of  a  blocking flow of $I(e)$,  and
 it holds $c(e) < \tilde{a}$.

\item[{\rm (ii)}]
One of the child-edges of $e$ is dominating if $c(e)<\tilde{a}$.
Edge $e=(u,v)$ satisfies condition {\rm (b)};
 if $vw_1$ or $vw_2$, say $vw_1$  is dominating, then
 there is a terminal $t^*\in T_{vw_1}$ such that
 $g({u,t^*})\geq \min\{\tilde{a}, c(e)\}$ holds
for any decomposition $g$ of  a  blocking flow of $I(e)$
and
   $P_{v,t^*}$   is a dominating path.

\item[{\rm (iii)}]
  For any  integers $x_1, x_2$ and $x$ such that $x_i\in \Psi(vw_i)$, $i=1,2$ and
  $x=x_1+x_2-2y$ for some integer $y\in [0, \min\{x_1,x_2\}]$,
let  $f_i$, $i=1,2$ be a blocking  flow of $I(vw_i)$ with  $f_i(vw_i)=x_i$.
Then $x\geq \tilde{a}$ holds.
 When  $\tilde{a}\leq c(e)$, any function $f=(x,f_1,f_2)$ with $x\leq c(e)$
   is a blocking  flow  of  $I(e)$.

\item[{\rm (iv)}]
 If $I(e)$ admits a blocking flow $f$ with $f(e)<c(e)$,
then $f(e)\in \langle \tilde{a}, \tilde{b} \rangle $.

\item[{\rm (v)}]
Assume that $c(e)<\tilde{a}$ and $vw_1$ is  dominating.
Let $P_{v,t^*}$ be the dominating path  in {\rm (iii)}
and let $\delta_e= \tilde{a}-c(e)$.
There is a blocking flow $f$ of $I(e)$ with $f(e)=c(e)$,
which can be constructed  as
\[ f=(c(e), f_1+(P_{v,t^*}, -\delta_e ), f_2) \]
by choosing
a  blocking flow  $f_1$  of   $I(vw_1)$  with $f_1(vw_1)=a(vw_1)$ and
a  blocking flow   $f_2$ of $I(vw_2)$ with $f_2(vw_2)=b(vw_2)$.

\item[{\rm (vi)}]  Edge $e=(u,v)$ satisfies condition {\rm (a)}; i.e.,
$\Psi(e)= ( \langle \tilde{a}, \tilde{b}\rangle   \cap   [0, c(e)]) \cup \{c(e)\}$.

\end{enumerate}
}

\pf{
 (i)  Let $f$  be  a blocking flow   of $I(e)$ such that
 $e\in E(V_f(t))$ for some terminal $t\in T_e$, where $c(e)=f(e)$.
Let $e_0,e_1,\ldots,e_p$   be the sequence of edges in $P_{u,t}$
such that
$e_i$ is the parent-edge of edges $e_{i+1}$ and $e'_{i+1}$
as shown  in Fig.~\ref{fig:lowest_flow_path}(a),
where   $e_0=e=(u,v)$ and  $e_p$ is the edge $e_t$ incident to the terminal $t$,
Note that $c(e_i)\geq 1+ f(e_i)$ for all $i=1,2,\ldots,p$ by definition of $V_f(t)$.

Since $f$ is blocking, set $V_f(t)$ is blocked by $f$ by definition,
and thereby
every positive-path $P_{s,s'}$ with $s,s'\in T_e\cup\{u\}$ of $f$  contains an edge in $E(V_f(t))$
 only when $t\in \{s,s'\}$
by \refl{integral-blocking-flow}(i).
This means that, for each $i=1,2,\ldots,p$,  $f(e_{i})= f(e_{i-1})+f(e'_{i})$,
from which $f(e_{i})= f(e_0)+\sum_{1\leq j\leq i}f(e'_{j})$.
 This proves that $g(u,t)\geq f(e_0)=c(e_0)$ for any  decomposition $g$ of $f$.

Since $V_f(t)$ is blocked by $f$, it holds $f(e_t) = f(V_f(t)) = c(V_f(t))-|\mathrm{odd}(V_f(t))|$, which implies that,
for each  $i=1,2,\ldots,p$,
\[  f(e'_i)=f(V_{e'_i}-V_f(t))= c(V_{e'_i}-V_f(t))-| \{ W \in \mathrm{odd}(V_f(t)) \mid W\subseteq V_{e'_i}\}| , \]
\[ f(e_W)=c(e_W)-1
\mbox{ for the parent-edge $e_W$ of each odd set
            $W\in \mathrm{odd}(V_f(t))$ with $W\subseteq V_{e'_i}$. } \]

\begin{figure}[t]
\begin{center}
\includegraphics[scale=0.40]{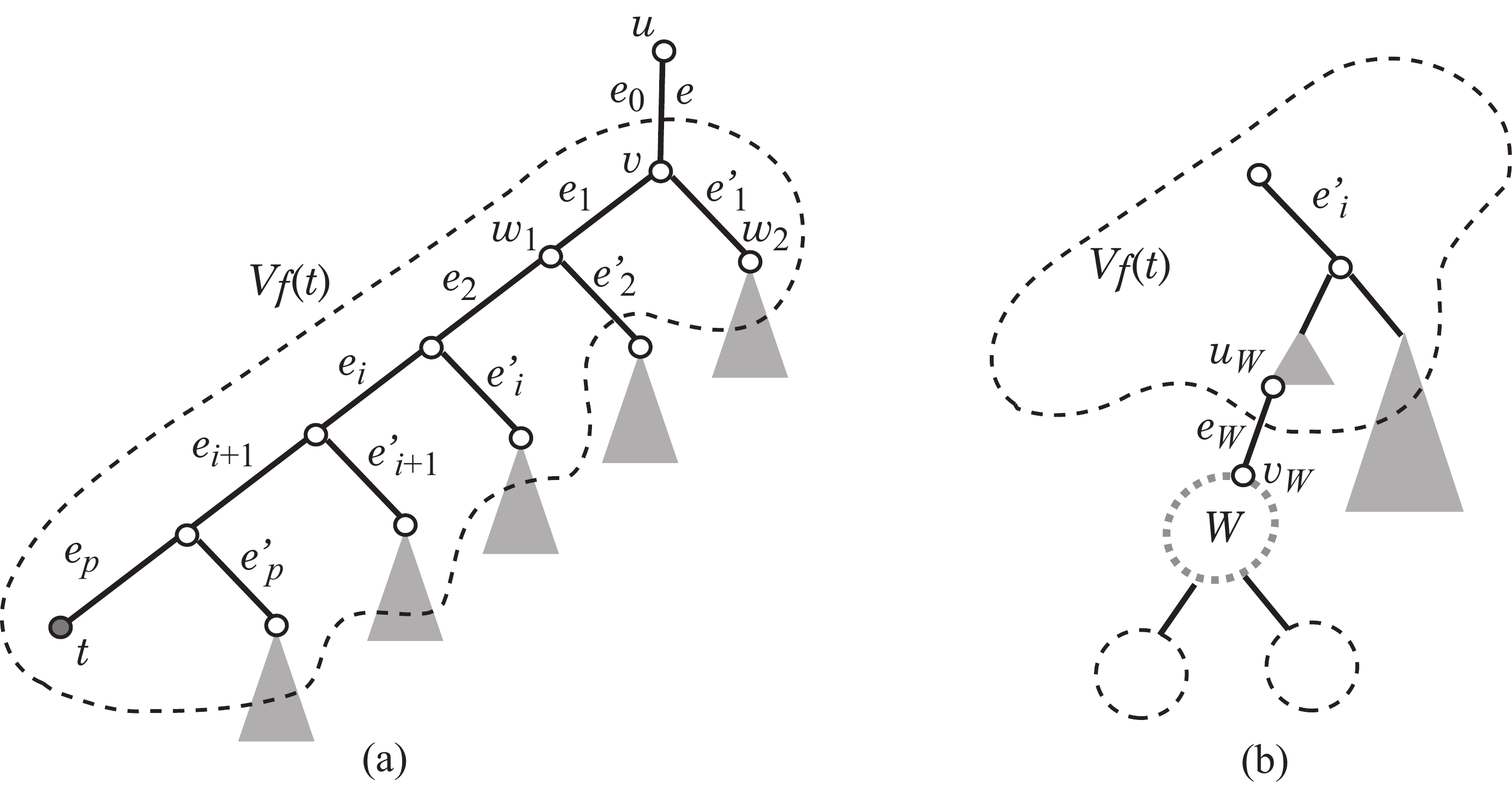}
\end{center}
\caption{(a) Terminal set $V_f(t)$ with $e\in E(V_f(t))$ and path $P_{u,t}$;
(b) an edge $e''_i$ with $h(e'_i)=b(e'_i)$ and an
odd set $W\in odd(V_f(t))$ with $W\subseteq V_{e'_i}$ . }
\label{fig:lowest_flow_path}
\end{figure}

First we prove that
$f(e'_i)\geq b(e'_i)$  for each $i=1,2,\ldots,p$.
For some $i$, assume indirectly that $f(e'_i) < b(e'_i)$.
Since $b(e'_i)\in \Psi(e'_i)$, the instance $I(e'_i)$  has
a blocking flow $h$ with $h(e'_i) = b(e'_i)$,
as shown  in Fig.~\ref{fig:lowest_flow_path}(b).
Since $f(V_{e'_i}-V_f(t))=f(e'_i) < b(e'_i) = h(e'_i) \leq h(V_{e'_i}-V_f(t))$ and
$f(e')=\min\{c(e'),c(e')-1\}$ for any edge $e'\in E(V_{e'_i}-V_f(t))$,
we see that  $V_{e'_i}$ must contain an odd set $W\in \mathrm{odd}(V_f(t))$
such that $c(e_W)-1 = f(e_W) < h(e_W)=c(e_W)$.
By applying \refl{integral-blocking-flow}(iv)
to the flow $h$ at
the saturated edge $e_W$, we have
 $c(e_W)=h(e_W)\in \Psi(e_W)$.
On the other hand,
by applying \refl{integral-blocking-flow}(v)  to $f$
at the parent edge $e_W$, we have
$c(e_W)-1=f(e_W)\in \Psi(e_W)$.
Hence  two consecutive integers $c(e_W)-1$ and $c(e_W)$ belong to $\Psi(e_W)$, which contradicts
 that $e_W$ satisfies $\Psi(e_W)=\langle a(e_W),b(e_W) \rangle$ by  (a).
Therefore
$f(e'_i)\geq b(e'_i)$  for each $i=1,2,\ldots, p$,
from which
\[c(e_{i}) \geq 1+ f(e_{i})\geq 1+ c(e_0)+\sum_{1\leq j\leq i}b(e'_{j}) \mbox{ for }
  i=1,2,\ldots,p. \]

Next for each $i=p,p-1,\ldots,1$,   we show that $e_i$ is dominating and derive a lower bound on $a(e_i)$.
For the leaf-edge $e_p$ incident to terminal $t$, it holds
\[ a(e_p)=b(e_p)=c(e_p). \]
Hence  $a(e_p)=c(e_p)\geq 1+ f(e_p) \geq  1+ c(e_0)+\sum_{1\leq j\leq p}b(e'_{j}) \geq 2+b(e'_p)$,
 and $e_p$ is dominating.
 By condition (a) for edge $e_{p-1}$,  we obtain
    $a(e_{p-1}) =  \min \{ \tilde{a}(e_{p-1})  , c(e_{p-1}) \} = \min \{a(e_{p}) - b(e'_{p})  , c(e_{p-1}) \}
         \geq \min \{1+c(e_0)+\sum_{1\leq j\leq p-1}b(e'_{j}) , c(e_{p-1}) \}
            = 1+ c(e_0)+\sum_{1\leq j\leq p-1}b(e'_{j}) $,
 since $c(e_{p-1})\geq 1+ c(e_0)+\sum_{0\leq j\leq p-1}b(e'_{j})$.
By applying condition (a) to $e_i$ repeatedly, we see that
\[a(e_{i})   \geq  1+ c(e_0)+\sum_{1\leq j\leq i}b(e'_{j}) \mbox{ for } i=1,2,\ldots,p-1,
\mbox{ and } \tilde{a} \geq 1+ c(e_0) ,\]
and edges $e_p,e_{p-1},\ldots,e_1$ are dominating edges.
Therefore  $c(e_0) <  \tilde{a}$ and $P_{v,t}$ is a dominating path. \\

\noindent   (ii)
If $c(e)<\tilde{a}$, where $\tilde{a}\geq c(e)+1\geq 2$, then
 one of  $vz_1$ and $vz_2$ is dominating,  since otherwise
    $\tilde{a}\in\{ 0,1\}$ would hold by applying (a) to $\Psi(vw_1)$ and $\Psi(vw_2)$.

 Assume that $vw_1$ is dominating, i.e., $a(vw_1)\geq b(vw_2)+2\geq 2$.
 Hence if $vw_1$ is a non-leaf-edge with two child-edges $w_1z_1$ and $w_1z_2$,
 then one of  $w_1z_1$ and $w_1z_2$ is dominating,  since otherwise
    $a(vw_1)\leq 1$ would hold by applying (a) to $\Psi(w_1z_1)$ and $\Psi(w_1z_2)$.
This implies that any dominating edge is a leaf-edge or has
a dominating child-edge, and hence
 there is a terminal $t^*\in T_{vw_1}$ such that
the path  $P_{v,t^*}$ from $u$ to $t^*$  is a dominating path .

Let $f$ be an arbitrary blocking flow in $I(e)$,
where $\mathcal{X}=\{V_f(t)\mid t\in T_e\}$ is a cut-system of $T_e$ blocked
by $f$ by definition.

First consider the case where  $e\in E(V_f(t))$ for some terminal $t\in T_e$.
By the result of (i),
$t=t^*$ must hold and  $g({u,t^*})\geq c(e)$ holds
for any decomposition $g$ of  $f$,
and we obtain $g({u,t^*})\geq \min\{\tilde{a}, c(e)\}$, as required.

Next assume that $e$ is not in $E(V_f(t))$ for any terminal $t\in T_e$.
Hence no cut in $\mathcal{X}$ contains any of the end vertices of $e$, and
$\mathcal{X}$ can be partitioned into
$\mathcal{X}_i=\{V_f(t)\mid t\in T_{vw_i}\}$, $i=1,2$.
In this case, for each $i=1,2$,
 the function $f_i$ induced from $f$ into $I(vw_i)$ is a blocking flow in $I(vw_i)$,
 since $\mathcal{X}_i$ is a cut-system  of $T_{vw_i}$ blocked by $f_i$.
To derive a contradiction, assume that there is a decomposition $g$ of  $f$
such that $g({u,t^*})< \min\{\tilde{a}, c(e)\}$, where  $\tilde{a}=a(vw_1)-b(vw_2)$.
Let $y$ be the amount of flows of $f$ that pass through $v$, i.e.,
  $y=\sum \{g(t,t') \mid t\in T_{vw_1}, ~t'\in T_{vw_2} \}$, where
we have
$x\leq f(vw_2)=f_2(vw_2)\leq  b(vw_2)$ since $f_2$ is a  blocking flow in $I(vw_2)$.
Based on $g$, we construct a decomposition $g_1$ of $f_1$ in $I(vw_1)$ as follows:
Recall that $v$ is a terminal in $I(vw_i)$, $i=1,2$.
$g_1(t,t')=g(t,t')$ for every two terminals $t,t'\in T_{vw_1}-\{t^*\}$ and
$g_1(t,v)=g(t,u)+\sum_{t'\in T_{vw_2}}g(t,t')$ for each terminal $t\in  T_{vw_1}$.
Since each path  in $G_e$ that contains an edge in $G_{vw_1}$ appears in one of the above two cases,
we see that $g_1$ is a decomposition of $f_1$ in $I(vw_1)$.
In particular,  $g_1(t^*,v)=g(t^*,u)+\sum_{t'\in T_{vw_2}}g(t^*,t') \leq  g(t^*,u)+y
  < (a(vw_1)-b(vw_2)) + b(vw_2) =a(vw_1)$.
  This, however, contradicts that condition for edge $vw_1$, where $g'(v,t^*)\geq a(vw_1)$ must hold
  for any decomposition $g'$ of a blocking flow in $I(vw_1)$.
 Therefore, there is no decomposition $g$ of  $f$
such that $g({u,t^*})< \min\{\tilde{a}, c(e)\}$.
  \\

\noindent   (iii)
Let $\mathcal{X}_i=\{V_{f_i}(t) \mid t\in T_{vw_i}\}$,
which is a cut-system of $T_{vw_i}$ blocked by $f_i$.
 Then the function $f=(x,f_1,f_2)$ is a flow in $I(e)$, since
 $f$ is obtained from $f_i$, $i=1,2$ as follows.
For   each, regard $f_i$ as a set of paths with unit flow values
 between terminals in $T_{vw_i}$, and let $\mathcal{P}^i_v$ denote
 the set of such paths that end with terminal $v$ in $f_i$.
 For each  for each $i=1,2$, choose   $x_i-y$ paths in $\mathcal{P}^i_v$ and
  extend them into paths that end at $u$.
 Then join the remaining $y$ paths in $\mathcal{P}^1_v$ and $y$ paths in $\mathcal{P}^2_v$
  pairwise to construct $y$ paths that join terminals in $T_{vw_1}$ and $T_{vw_2}$.
 That is how a flow $f$ in $I(e)$  with $f(e)= (x_1-y)+(x_2-y)$ is constructed,
 where $f$ is feasible if $c(e)\geq x$.

By construction of $x$ from $x_i\in \Psi(vw_i)$, $i=1,2$,
it holds $x\in  \langle \tilde{a}, \tilde{b} \rangle $, and
we have $x\geq \tilde{a}$.

We easily see that $\{V_f(t)\mid t\in T_e\}$ is equal to $\mathcal{X}_1\cup \mathcal{X}_2$
and is a cut-system of $T_e$ blocked by the function $f=(x,f_1,f_2)$.
Then if $\tilde{a}\leq c(e)$, any
  function $f=(x,f_1,f_2)$ with $x\leq c(e)$
 is a blocking  flow  of  $I(e)$. \\

\noindent   (iv)
Assume that $I(e)$ admits a blocking flow $f$ with $f(e)<c(e)$,
and let $g$ be  a decomposition of $f$.
Then $\mathcal{X}=\{V_f(t)\mid t\in T_{e}\}$ is a cut-system
of $T_e$ blocked by $f$ by definition.
Since  $f(e)<c(e)$,
edge $e$ is not saturated by $f$ and  is not contained in $E(V_f(t))$
of any terminal $t\in T_e$,
and
for each $i=1,2$,
  $\mathcal{X}_i=\{V_f(t)\mid t\in T_{vw_i}\}$ is a cut-system of $T_{vw_i}$ blocked by
the function $f_i$ in $I(vw_i)$ induced from $f$.
Hence $f_i$ is a blocking flow in $I(vw_i)$ and
$f_i(vw_i)\in \Psi(vw_i)=\langle a(vw_i), b(vw_i) \rangle$ by condition (a).
Let $y$ be the amount of flows of $f$ that pass through $v$, i.e.,
  $y=\sum \{g(t,t') \mid t\in T_{vw_1}, ~t'\in T_{vw_2} \}$,
  where  $y\in [0, \min\{f_1(vw_1),f_2(vw_2)\}]$ and
 $f(e)=  f_1(vw_1)+f_2(vw_2) -2y$.
 Hence
  $f(e)$ satisfies the condition of elements in
   $\langle a(vw_1), b(vw_1) \rangle \otimes \langle a(vw_2), b(vw_2) \rangle
  =\langle \tilde{a}, \tilde{b} \rangle $, i.e., $f(e)\in \langle \tilde{a}, \tilde{b} \rangle$. \\

\noindent   (v)
Assume that $\tilde{a}> c(e)$, where $\tilde{a}\geq c(e)+1\geq 2$.
For a  blocking flow $f_i$  of   $I(vw_i)$, $i=1,2$ such that
  $f_1(vw_1)=a(vw_1)$ and   $f_2(vw_2)=b(vw_2)$,
   let $g_i$, $i=1,2$, be a decomposition of $f_i$.
Condition (b) with edge $(v,w_1)$ implies $g_1(v,t^*)\geq a(vw_1)$,
from which $g_1(v,t^*)= a(vw_1)$ since $g_1(v,t^*)\leq f_1(vw_1)=a(vw_1)$.

 Based on $g_i$, $i=1,2$, we construct flows $f_{1,2}$ and $f$ in $I(e)$ and
 flows $f'_1$ and $f''_1$ in $I(vw_1)$ and their decompositions
 $g_{1,2}$, $g$, $g'_1$ and $g''_1$ as follows.

 Let $f_{1,2}=(\tilde{a}, f_1, f_2)$ be a function in $I(e)$.
 A decomposition $g_{1,2}$ of $f_{1,2}$ can be obtained by
 $g_{1,2}(t^*,t)=g_2(v,t)$ for each terminal $t\in T_{vw_2}$,
 $g_{1,2}(t^*,u)=g_1(t^*,v)-\sum_{t\in T_{vw_2}}g_2(v,t) = a(vw_1)-b(vw_2)=\tilde{a}$, and
 $g_{1,2}(s,s')=g_i(s,s')$ for any other terminal pairs $s,s'\in T_{vw_i}$, $i=1,2$.
Then   $f_{1,2}$ is a flow in $I(e)$ but  not feasible
 since $f_{1,2}(e)=\tilde{a}> c(e)$.

 By decreasing the value of $g_1(t^*,v)$ by $c(e)-\tilde{a}$,
 we obtain a flow $f'_1=f_1+(P_{v,t^*}, -(\tilde{a}-c(e)))$ in $I(vw_1)$.
 A decomposition $g'_1$ of  $f'_1$ is given by
 $g'_1(t^*,v) =g_1(t^*,v) - ( c(e)-\tilde{a} )$ and
  $g'_1(s,s')=g_1(s,s')$ for  any other terminal pairs $t,t'\in T_{vw_1}$.
  By decreasing the value of
    $g_{1,2}(t^*,u)$ by $c(e)-\tilde{a}$,
    we obtain  a flow $f=(c(e), f'_1, f_2)$  in $I(e)$, which is feasible
     since  $g_{1,2}(t^*,u)= \tilde{a}>c(e)$.
 A decomposition  $g$ of $f$ is given by
 $g(t^*,u)=g_{1,2}(t^*,u) - ( c(e)-\tilde{a} )$ and
  $g(s,s')=g_{1,2}(s,s')$ for  any other terminal pairs $s,s'\in T_{e}$.

In the rest of the proof, we show that  $f=(c(e),f'_1,f_2)$ is  a blocking flow in $I(e)$.
For this, it suffices to show that
$V_f(t)\cap V_f(t')=\emptyset$ for any two terminals $t,t'\in T_e$
and $f(e_t)=f(V_f(t))$ for all terminals $t\in T_e$ by \refl{integral-blocking-flow}(iii).
For each $i=1,2$,
let $\mathcal{X}_i=\{V_{f_i}(t)\mid t\in T_{vw_i}\}$,
which is a cut-system of $T_{vw_i}$ blocked by $f_i$
by definition.
Then $\{V_f(t)\mid t\in T_e\}=  (\mathcal{X}_1-\{V_{f_1}(t^*)\} )\cup\{V_f(t^*)\}\cup \mathcal{X}_2$.
Hence it suffices to prove that (1)
  $V_f(t^*)$ is disjoint with any other cut in $\mathcal{X}_1\cup \mathcal{X}_2$;
  and
  (2) $f(e_{t^*})=f(V_f(t^*))$.

We first prove (1).
Let $V_{f}(t)$ be a terminal  set  in $\mathcal{X}_1\cup \mathcal{X}_2$ with $t\neq t^*$.
Since $g_1(v,t^*)\geq a(vw_1)\geq 2$ along the dominating path  $P_{v,t^*}$
by condition (b) for edge $vw_1$,
set $V_{f}(t)$  is disjoint with $P_{v,t^*}$  by \refl{integral-blocking-flow}(i).
On the other hand, no edge in $P_{v,t^*}$ is saturated,
the set $V_f(t^*)$ contains all the vertices in $P_{v,t^*}$.
Since the parent-edge of $V_{f}(t)$ is saturated by $f_i$ with $i\in \{1,2\}$,
the set $V_f(t^*)$ is disjoint with $V_{f}(t)$, as required.

We next prove (2).
To derive a contradiction, we assume that there are terminals $t,t'\in T_e-\{t^*\}$
such that $g(t,t')\geq 1$ for the above decomposition $g$ of $f$
and
path $P_{t,t'}$ is not disjoint with $V_f(t^*)$,
i.e.,
$V_f(t^*)$ contains the least common ancestor $\ell$ of $t$ and $t'$.
Let $\ell'$ be  the least common ancestor  of $\ell$ and $t^*$.
Recall that $g(s,s')=0$ for any $s\in T_{vw_1}-\{t^*\}$ and $s'\in T_{vw_2}$
by construction of $g$ from $g_1$ and $g_2$.
Hence $t,t'\in T_{vw_1}-\{t^*\}$ and $\ell\neq u\neq \ell'$.
We modify $f_1$ into a  function
$f''_1:=f_1+(P_{t,t'},-1)+(P_{v,t^*},-1) +(P_{v,t},1) +(P_{t^*,t},1)$,
which is clearly a feasible flow in $I(vw_1)$,
and a decomposition $g''_1$ of $f''_1$ can be obtained
by setting
 $g''_1(t,t')=g_1(t,t')-1$,
 $g''_1(v,t^*)=g_1(v,t^*)-1$,
 $g''_1(v,t)=g_1(v,t)+1$,
 $g''_1(t^*,t)=g_1(t^*,t)+1$,
 and $g''_1(s,s')=g_1(s,s')$
for any other pair of terminals $s,s'\in T_{vw_1}$.

We prove that $\mathcal{X}_1$ is still blocked by $f''_1$.
To show that each set $V_{f_1}(t)\in \mathcal{X}_1$ is blocked by $f''_1$,
it suffices to prove that $V_{f_1}(t) =V_{f''_1}(t)$ and
$f''_1(e')=f_1(e')$ for each edge $e'\in E(V_{f_1}(t))$.
We see that
$f''_1(e')<f_1(e')$ can hold only when $e'$ is on the path $P_{v,t^*}$
and $f''_1(e')>f_1(e')$ can hold only when
$\ell$ is not on the path  $P_{v,t^*}$ and
$e'$ is on the path between $\ell$ and $\ell'$.
Hence $f''_1(e')\neq f_1(e')$ holds only for an edge $e'$ in the graph induced from $G$ by $V_f(t^*)$.
Since $V_f(t^*)$ is disjoint with any  set $V_{f_1}(t)\in \mathcal{X}_1$ with $t\neq t^*$,
we see that $V_{f_1}(t) =V_{f''_1}(t)$ and $f''_1(e')=f_1(e')$ for all $e'\in E(V_{f_1}(t))$.
If $\ell\in V_{f_1}(t^*)$, then the positive-path $P_{t,t'}$ would not be disjoint with $V_{f_1}(t^*)$,
contradicting  \refl{integral-blocking-flow}(i).
Hence $\ell\not\in V_{f_1}(t^*)$.
If $\ell'\in V_{f_1}(t^*)$, then $\ell'$ and $\ell \in V_f(t^*) - V_{f_1}(t^*)$ is connected
by a path $P_{\ell',\ell}$ such that $c(e')-f_1(e')\geq 2$ for all edges $e'\in E(P_{\ell',\ell})$,
contradicting that $c(e')-f(e')\leq 1$ for all edges $e'\in E(V_{f_1}(t^*))$.
Hence  $V_{f_1}(t) =V_{f''_1}(t)$ and $f''_1(V_{f_1}(t))=f_1(V_{f''_1}(t))$, implying that
$V_{f_1}(t)\in \mathcal{X}_1$ is blocked by $f''_1$.

Therefore $\mathcal{X}_1$ is  blocked by $f''_1$,  and
$f''_1$ is a blocking flow in $I(vw_1)$ with  $g''_1(v,t^*)=g_1(v,t^*)-1<a(vw_1)$.
This, however, contradicts that (b) holds for the blocking flow $f''_1$,
proving that (2) holds.

 From (1) and (2),  $\mathcal{X}$  is blocked by $f$,
 and   $f$ is a blocking flow in $I(e)$. \\

\noindent   (vi)
We distinguish two cases.
First assume that   $\tilde{a}\leq c(e)$.
 Then $\Psi(e)\supseteq  \langle \tilde{a}, \tilde{b}\rangle    \cap   [0, c(e)]$ by (ii) and
$\Psi(e)\subseteq \langle \tilde{a}, \tilde{b}\rangle  \cap   [0, c(e)]$   by (iv).
Next assume that $c(e)<\tilde{a}$.
 Then $\Psi(e)\subseteq \{c(e)\}$ by (iv) and
$\Psi(e)\supseteq \{c(e)\}$ by (v).
}

 \section{Algorithm Description}  \label{sec:algorithm}
   Based on \refl{induction_integral_flow},
   this section  gives a description of a linear-time algorithm
    for computing the  representations of flow values of blocking flows
    and constructing a maximum flow from the representations.

By \refl{induction_integral_flow}(ii) and (iv), we see by induction
that every edge in $E$ satisfies conditions (a) and (b).
By \refl{induction_integral_flow}(iii) and (v), we know
  how to construct a blocking flow in $I(e)$ for some edge $e$
  from blocking flows in $I(e_1)$ and $I(e_2)$ of the child-edges $e_1$ and $e_2$ of $e$.
  By \refl{last_integral_flow}, it suffices to construct a blocking flow in $I=I(e_r)$
  with $f(e_r)=b(e_r)$.
  For this, we first compute    the integers
$\tilde{a}(e)$, $\tilde{b}(e)$,   $a(e)$ and $b(e)$ for each edge  $e\in E$
  according to  (\ref{eq:update-1})  and   (\ref{eq:update-2})
  selecting edges in $E$ in a non-increasing order of depth,
  and identify all the dominating edges in $E$.
Next we apply  \refl{induction_integral_flow}(iii) and (v) repeatedly from edge $e_r$
to descendants of the edge in a top-down manner
to construct  a blocking flow in $I=I(e_r)$
  with $f(e_r)=b(e_r)$.
To implement the algorithm to run in linear time, we avoid reducing flow values repeatedly
along part of a dominating path.
We let
  $\sigma(e)$ to store the total amount of decrements  over each dominating edge $e$,
  i.e., $\sigma(e)$ is the summation of $\delta_{e'}$ in  \refl{induction_integral_flow}(v)
  over all dominating edges $e'$ that are
   ancestors  of $e$.
  An entire algorithm is given by the following   compact and succinct description.
\bigskip

\begin{figure}[htbp]
\begin{center}
\includegraphics[scale=0.40]{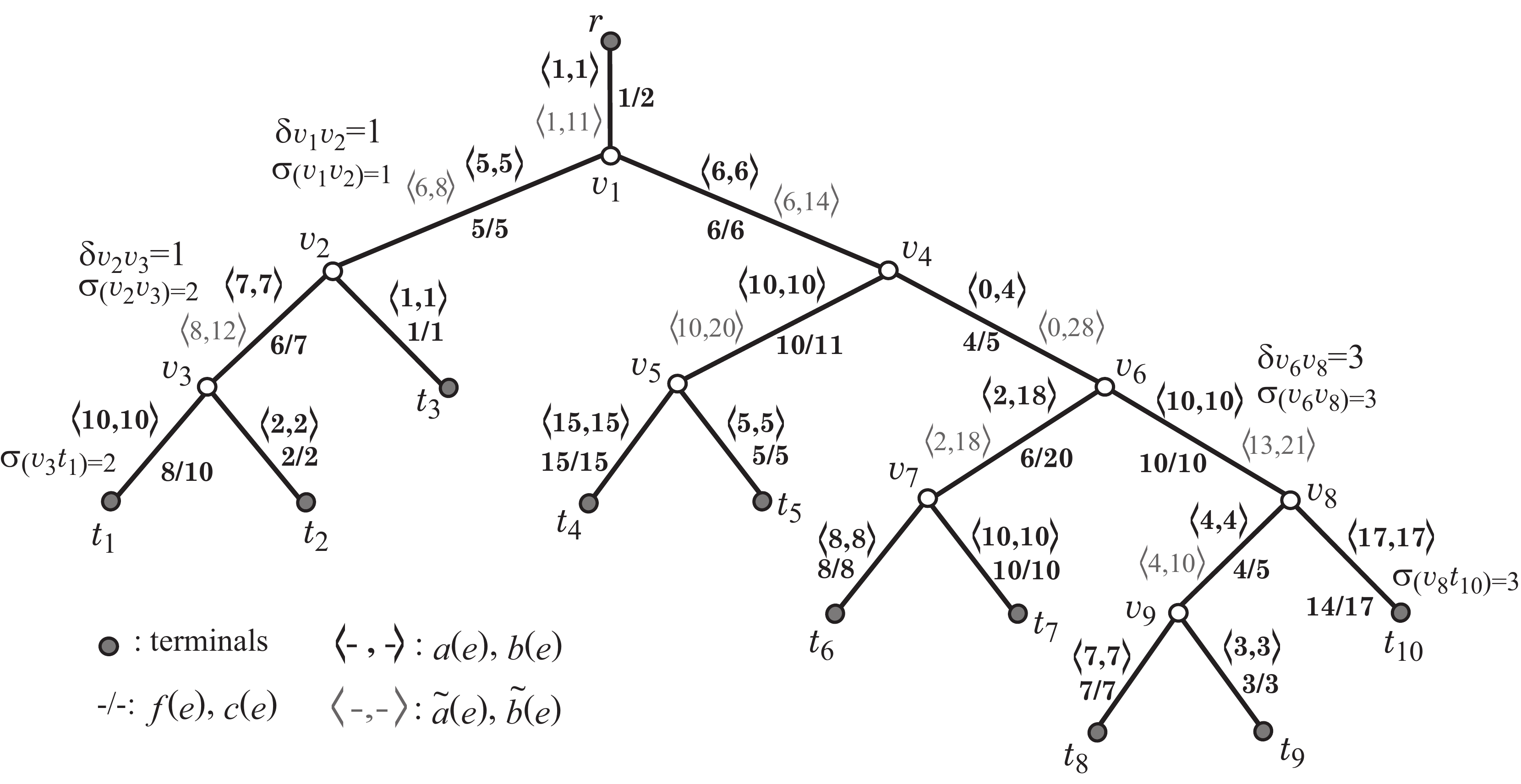}
\end{center}
\caption{A blocking flow $f$  with $f(e_r)=b(e_r)$ in the instance $I$ in Fig.~\ref{fig:instance_integer1} such that
 $2\alpha(f)=\sum_{t\in T}f(e_t)=1+8+2+1+15+5+8+10+7+3+14=74$,
where the pair of flow value $f(e)$ and capacity $c(e)$ for each edge is indicated
by $f/c$ beside the line segment for edge $e$.
The non-zero values for $\delta_e$ and $\sigma(e)$ are indicated beside the corresponding edge $e$.  }
\label{fig:instance_integer3}
\end{figure}

\begin{algorithm}
\caption{\textsc{BlockFlow}}
\begin{algorithmic}[0]

\REQUIRE An instance $I=(G=(V,E),T,c)$ rooted at a terminal $r\in T$. \\
\ENSURE A maximum flow $f$ in $I$. \\
\smallskip
Compute   the integers
$\tilde{a}(e)$, $\tilde{b}(e)$,   $a(e)$ and $b(e)$ for each edge  $e\in E$
  according to  (\ref{eq:update-1})  and   (\ref{eq:update-2})
  selecting edges in $E$ in a non-increasing order of depth; \\
\smallskip
 $x(e_r):=b(e_r)$; $\sigma(e_r):=0$; \\

\FOR{each edge $e\in E$ selected in a non-decreasing order of depth}
\STATE
  $f(e):=x(e)-\sigma(e)$; \\
  \IF{$e$ is not a leaf edge}
  \STATE
    /* Denote by $e_1$ and $e_2$ the child-edges   of $e$ */
    \IF{$\tilde{a}(e)\leq c(e)$}
    \STATE  Choose integers $x_1\in  \langle  a(e_1), b(e_1) \rangle$ and
    $x_2\in \langle  a(e_2), b(e_2) \rangle$   such that  \\
       $x(e) =x_1+x_2-2y$ for some integer  and $y\in [0, \min\{x_1,x_2\}]$; \\
 $x(e_1)=x_1$;~  $x(e_2)=x_2$; \\
\IF{$e_i$ is dominating for $i=1$ or 2}
\STATE
      $\sigma(e_i):=\sigma(e)$ and $\sigma(e_j):=0$ for $j\in \{1,2\}-\{i\}$
\ELSE
\STATE $\sigma(e_1):= \sigma(e_2):=0$ \\
\ENDIF
\ELSE
\STATE
    /*   $c(e_0)< \tilde{a}(e_0)$, where
    $e_0$ is dominating, and exactly one of $e_1$ and $e_2$  is\\
     dominating;  assume that $e_1$   is dominating without loss of generality. */ \\
     $x(e_1)=a(e_1)$;~   $x(e_2)=b(e_2)$;~  $\delta_{e_1}:= a(e_1)-c(e)$;\\
    $\sigma(e_1):= \sigma(e)+\delta_{e_1}$;~  $\sigma(e_2):= 0$ \\
\ENDIF
\ENDIF
\ENDFOR
\end{algorithmic}
\end{algorithm}

\begin{figure}[htbp]
\begin{center}
\includegraphics[scale=0.40]{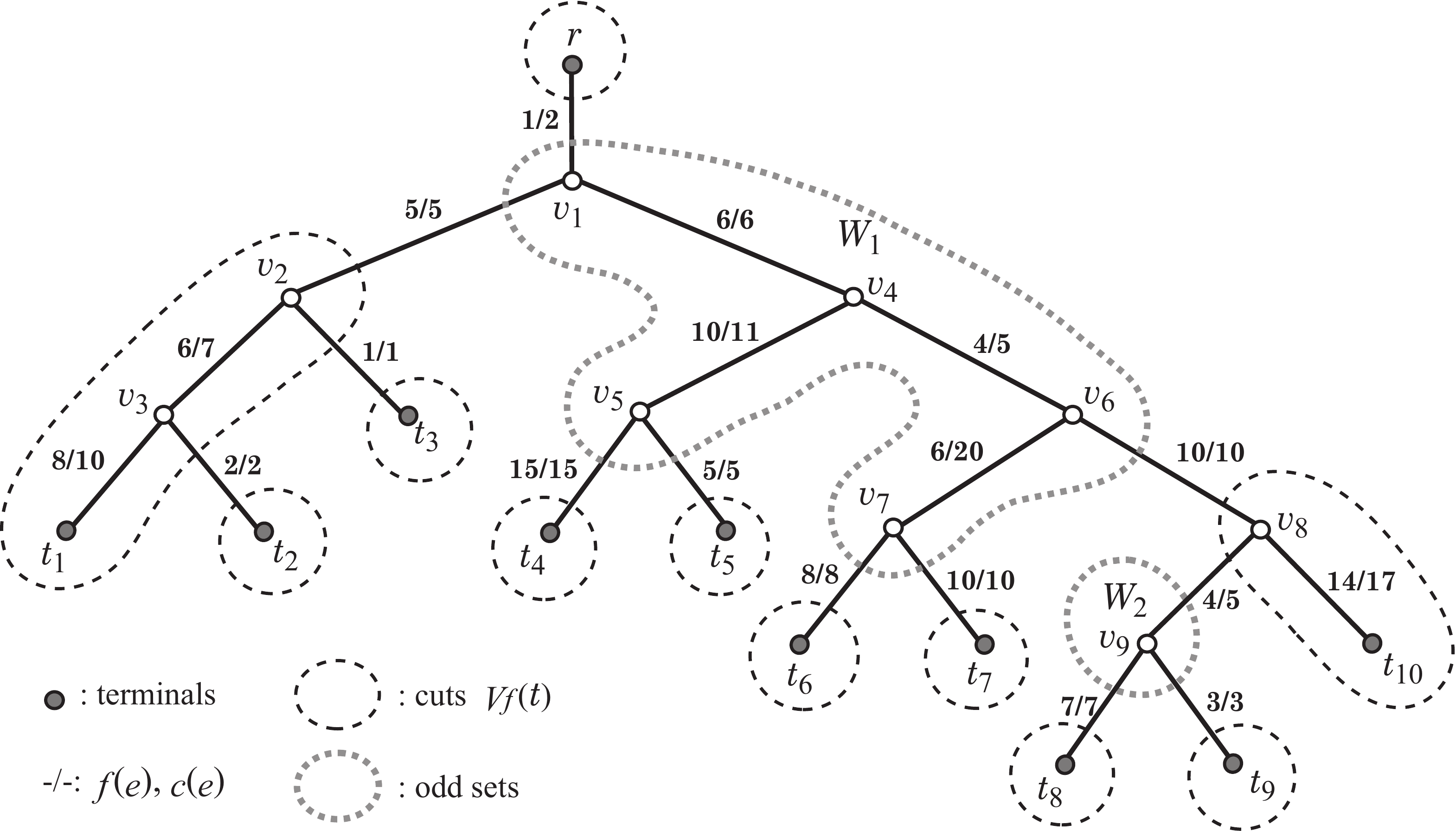}
\end{center}
\caption{The cut-system $\mathcal{X}=\{V_f(t)\mid t\in T\}$ for the blocking flow $f$ in  Fig.~\ref{fig:instance_integer3}, where the set $V-\cup_{X\in \mathcal{X}}  X$ induces from $G$
two odd sets $W_1\in \mathrm{odd}(V_f(r))$ and $W_2\in \mathrm{odd}(V_f(t_{10}))$,
and it holds that
$\gamma(\mathcal{X})-\kappa(\mathcal{X})
= \sum_{t\in T}c(V_f(t))-2=  2+(2+1+5)+2+1+15+5+8+10+7+3+(5+10)- 2
    =   74$. }
\label{fig:instance_integer4}
\end{figure}

The algorithm runs in linear time, because   it executes
an $O(1)$-time procedure to each edge in $E$ in constant time.
Fig.~\ref{fig:instance_integer3} illustrates a result obtained from
the instance $I$ in Fig.~\ref{fig:instance_integer1}
by applying the algorithm.

After a maximum flow $f$ is constructed,
a minimum cut-system $\mathcal{X}$ to a given instance
can be constructed in linear time by \refl{last_integral_flow}.
Fig.~\ref{fig:instance_integer4} illustrates
 the cut-system $\mathcal{X}=\{V_f(t)\mid t\in T\}$
  for the blocking flow $f$ in  Fig.~\ref{fig:instance_integer3}, which indicates that the flow $f$ is maximum
 because $2\alpha(f)=\sum_{t\in T}f(e_t)=74=\gamma(\mathcal{X})-\kappa(\mathcal{X})$
 holds.

From the above argument, the next theorem is established.

\begin{theorem}\label{th:main}
Given a tree instance $(G,T,c)$,
a feasible integral multiflow $f$
and a cut-system ${\cal X}$ with
$\alpha(f)=(\gamma({\cal X})-\kappa({\cal X}))/2$
can be found in $O(n)$ time and space,
where  $f$ is a maximum integral multiflow.
\end{theorem}

\section{Concluding Remarks}\label{conclusions}

In this paper, we revealed a recursive formula among flow values of blocking flows in rooted instances
and designed a linear-time  dynamic programming algorithm for computing
a maximum integral flow in a tree instance.
The optimality of flows is ensured by the property of the formula, by which
we can always construct the corresponding dual object, i.e., a minimum cut-system
that satisfies (\ref{ingeter-min-max}) by equality.

It would be interesting to characterize similar recursive properties and design fast algorithms
for the maximum integral multiterminal flows in more general classes of graphs.



\begin{thebibliography}{10}

\bibitem{Cherkasskii}
B.V. Cherkasskii.
\newblock Reshenie odnoi zadachi o mnogoproduktovykh potokakh v seti [russian;
  a solution of a problem of multicommodity flows in a network].
\newblock {\em \`{E}konomika i Matematicheskie Metody}, 13(1):143--151, 1977.

\bibitem{Costa:Survey}
Marie-Christine Costa, Lucas L{\'e}tocart, and Fr{\'e}d{\'e}ric Roupin.
\newblock Minimal multicut and maximal integer multiflow: a survey.
\newblock {\em European Journal of Operational Research}, 162(1):55--69, 2005.

\bibitem{Costa:tree}
Mariechristine Costa and Alain Billionnet.
\newblock Multiway cut and integer flow problems in trees.
\newblock {\em Electronic Notes in Discrete Mathematics}, 17(20):105--109,
  2004.

\bibitem{Cunningham:multiterminal}
William~H Cunningham.
\newblock The optimal multiterminal cut problem.
\newblock {\em {DIMACS} series in discrete mathematics and theoretical computer
  science}, 5:105--120, 1991.

\bibitem{FordFulkerson:flow}
J.~R. Ford and D.~R. Fulkerson.
\newblock {\em Flows in networks}.
\newblock Princeton university press, 1962.

\bibitem{FlowMulticut:approximateTheorems}
Naveen Garg, Vijay~V. Vazirani, and Mihalis Yannakakis.
\newblock Approximate max-flow min-(multi) cut theorems and their applications.
\newblock {\em {SIAM} Journal on Computing}, 25(2):235--251, 1996.

\bibitem{Garg:multicutinTrees}
Naveen Garg, Vijay~V. Vazirani, and Mihalis Yannakakis.
\newblock Primal-dual approximation algorithms for integral flow and multicut
  in trees.
\newblock {\em Algorithmica}, 18(1):3--20, 1997.

\bibitem{HKNR:treewidthflow}
Torben Hagerup, Jyrki Katajainen, Naomi Nishimura, and Prabhakar Ragde.
\newblock Characterizing multiterminal flow networks and computing flows in
  networks of small treewidth.
\newblock {\em Journal of Computer and System Sciences}, 57(3):366--375, 1998.

\bibitem{IKN:inner_euler}
Toshihide Ibaraki, Alexander~V. Karzanov, and Hiroshi Nagamochi.
\newblock A fast algorithm for finding a maximum free multiflow in an inner
  eulerian network and some generalizations.
\newblock {\em Combinatorica}, 18(1):61--83, 1998.

\bibitem{mader1978}
Wolfgang Mader.
\newblock {\"U}ber die {M}aximalzahl kantendisjunkter {A}-{W}ege.
\newblock {\em Archiv der Mathematik}, 30(1):325--336, 1978.

\end{thebibliography}
\end{document}